\documentclass[a4paper]{article}

\usepackage{INTERSPEECH2020}
\usepackage{multirow}
\usepackage{algorithm}  
\usepackage{algorithmic}

\title{
Incorporating VAD into ASR System by Multi-task Learning}
\name{Meng Li$^1$, Yan Xia$^1$, Feng Lin$^1$}
\address{
  $^1$Megatronix(Beijing) Technology Co., Ltd.}
\email{\{meng.li,yan.xia, feng.lin\}@megatronix.co}

\begin{document}

\maketitle
\begin{abstract}
  
When we use End-to-end automatic speech recognition (E2E-ASR) system for real-world applications, a voice activity detection (VAD) system is usually needed to improve the performance and to reduce the computational cost by discarding non-speech parts in the audio. Usually ASR and VAD systems are trained and utilized independently to each other. In this paper, we present a novel multi-task learning (MTL) framework that incorporates VAD into the ASR system. The proposed system learns ASR and VAD jointly in the training stage. With the assistance of VAD, the ASR performance improves as its connectionist temporal classification (CTC) loss function can leverage the VAD alignment information. In the inference stage, the proposed system removes non-speech parts at low computational cost and recognizes speech parts with high robustness. Experimental results on segmented speech data show that by utilizing VAD information, the proposed method outperforms the baseline ASR system on both English and Chinese datasets. On unsegmented speech data, we find that the system outperforms the ASR systems that build an extra GMM-based or DNN-based voice activity detector. 

\end{abstract}

\section{Introduction}

In recent years, there has been a growing trend in probing into end-to-end automatic speech recognition (E2E-ASR) system, which directly maps audio waves into text. The most popular E2E-ASR approaches include the connectionist temporal classification (CTC) \cite{Graves2006ConnectionistTC,Graves2014TowardsES} , the recurrent neural network transducer (RNN-T) \cite{Graves2012SequenceTW,Rao2017ExploringAD} and attention-based encoder-decoder architectures \cite{Bahdanau2015NeuralMT}. E2E-ASR models
show advantages over traditional methods in simplicity and outperform conventional ASR systems when trained on enough training data \cite{2017State}. However, most of the methods are based on the assumption that the input audio has been processed into short speech segments. The mismatch between the assumption and the real-world scene makes it difficult to transcribe unsegmented long audios and to recognize speech in real time by directly using E2E-ASR systems. To approach the problem, a voice activity detection (VAD) \cite{Atal1976APR,Ramirez2007VoiceAD} system is often built to detect the speech segments and to discard the non-speech segments in the input audio.

A number of techniques can be used for VAD. Unsupervised approaches include building VAD systems based on energy \cite{Woo2000RobustVA}, zero crossing rate \cite{Junqua1991ASO}, the periodicity measure \cite{Tucker1992VoiceAD}. Supervised VAD systems include support vector machines \cite{Mesgarani2006DiscriminationOS}, Gaussian mixture models (GMM) \cite{Lee2004NoiseRR}, deep neural networks (DNN) \cite{Zhang2013DeepBN,Ryant2013SpeechAD,Hughes2013RecurrentNN}. In recent years, DNN-based VAD systems have attracted much attention because they can extract more information from the input feature and achieve better performance than conventional VAD systems. However, it takes extra effort and additional memory resources to build a DNN-based VAD model. A viable solution is integrating ASR and VAD into one model.

Recently, attempts have been made to integrate ASR and VAD into an E2E Neural Network. In Yoshimura et al.'s study \cite{Yoshimura2020EndtoEndAS} , VAD is integrated into a CTC-based E2E-ASR model, in which blank labels from the CTC softmax output are regarded as the non-speech region. In Tao and Busso's study \cite{Tao2021EndtoEndAS}, a multi-task learning (MTL) framework, which has two classification layers on the top of the network, is proposed to perform both audiovisual ASR and audiovisual VAD tasks. In the above-mentioned studies, VAD and ASR share the same network architecture. However, compared with ASR, VAD is less complicated and needs fewer computational resources, which means that using the same model structure for VAD and ASR is not computationally efficient. 

In this work, we attempt to incorporate VAD into an E2E ASR system by leveraging multi-task learning approach. In the training stage, the model is firstly pre-trained with wav2vec 2.0 \cite{Baevski2020wav2vec2A} framework, a self-supervised framework for speech representation learning, which has shown its advantage of helping convergence and improving the performance in ASR. In this paper, we build the model based on wav2vec 2.0 to get high performance in both ASR and VAD tasks. Then, ASR and VAD tasks are jointly trained with a MTL technique. With the help of MTL, the model learns representations that are discriminative for all tasks and obtains the better generalization than models trained by single task learning (STL). Experimental results show that our MTL approach outperforms STL approach in both ASR and VAD tasks. 
To reduce the computational cost of VAD, we only use the output features from the bottom feature extraction module of the network architecture to perform the VAD task. This design is more consistent with human cognition process and infants' language learning process: In Jusczyk's study \cite{Jusczyk1999HowIB}, infants firstly have the ability to detect the language sound patterns before they recognize words. 
And to help the ASR system make better use of the information learned from VAD task, we propose a cross-task attention module to learn interactive information between ASR and VAD.
To support online speech recognition, we use a chunk-hopping mechanism, which enables the model to encode on segmented frame chunks one after another \cite{Dong2019SelfattentionAA}. And to eliminate unnecessary computational cost in the inference stage, we propose an online VAD\&ASR inference algorithm with high robustness. Experimental results show that when we transcribe unsegmented long audios with the online VAD\&ASR joint inference algorithm, the performance is nearly as good as what is achieved by transcribing short speech segments processed through human efforts. 

\section{Proposed Methods}

In this section, we will introduce our proposed approach, which combines VAD with ASR through MTL technique and is pre-trained with wav2vec 2.0 framework \cite{Baevski2020wav2vec2A}. 

\subsection{Architecture}
\label{sec:architecture}
\begin{figure}[t]
	\centering
	\includegraphics[width=\linewidth]{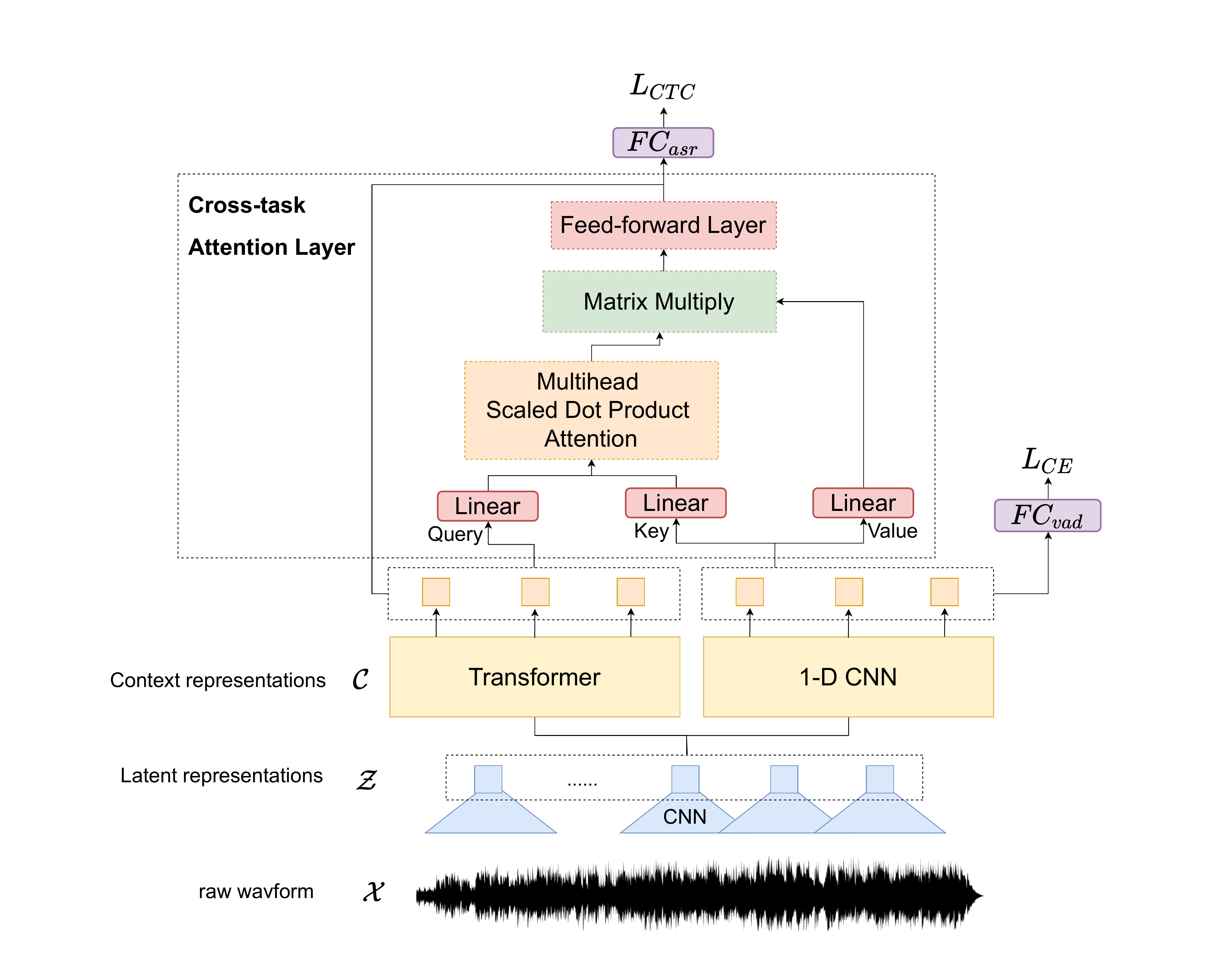}
	\caption{Proposed model architecture.}
	\label{fig:architecture}
\end{figure}

As is shown in Figure~\ref{fig:architecture}, the proposed model architecture is built on the base structure of wav2vec 2.0, which is composed of two parts: a multi-layer convolutional feature encoder and a context network which follows the Transformer architecture \cite{Vaswani2017AttentionIA, Devlin2019BERTPO}. The convolutional feature encoder maps the raw audio input $\mathcal{X}$ into latent representations, $\mathcal{Z}=(z_1,...,z_T)$. Then the context network takes the latent representations as input to build contextualized representations, $\mathcal{C}=(c_1,...,c_T)$. 

Based on the architecture of wav2vec2.0 , we perform ASR and VAD on different layers. For VAD task, to provide temporal information, we add a single group 1-D convolutional layer that is applied directly on the output embeddings of the feature encoder. Then we add an FC layer, which takes latent representations as input and outputs class representations of speech and non-speech, on the 1-D convolutional layer.

For ASR task, to better utilize the alignment information learned from VAD, we propose adding an Cross-task Attention Layer after the context network (Transformer). 

Derivate from Self-Attention\cite{Vaswani2017AttentionIA}, The Cross-task Attention Layer has three inputs: a pair of Key-Value vectors learned from one task(VAD) and a Query vector learned from another task(ASR). Then we let the Query vector learned from ASR attend to the entire sequence embedding learned from VAD via the Multihead Self-Attention mechanism. Finally, we add the ouput of the Cross-task Attention Layer to contextualized vector $c_t$ and send these through a prediction head.

As depicated in Figure~\ref{fig:architecture}, The Cross-task Attention Layer extracts features $\mathcal{G}=(g_1,...,g_T)$ from $Q$, $K$ and $V$ as follows: 
\begin{equation}
g_t = c_t + f_{g}(Q_{asr},K_{vad}, V_{vad}|\theta_g)
\label{eq0}
\end{equation}
where $f_g$ is the function played by Cross-task attention layer with parameter set $\theta_g$. 

\subsection{Chunk-hopping Mechanism}
\label{sec:chunk-hop}
\begin{figure}[t]
	\centering
	\includegraphics[height=4cm,width=7cm]{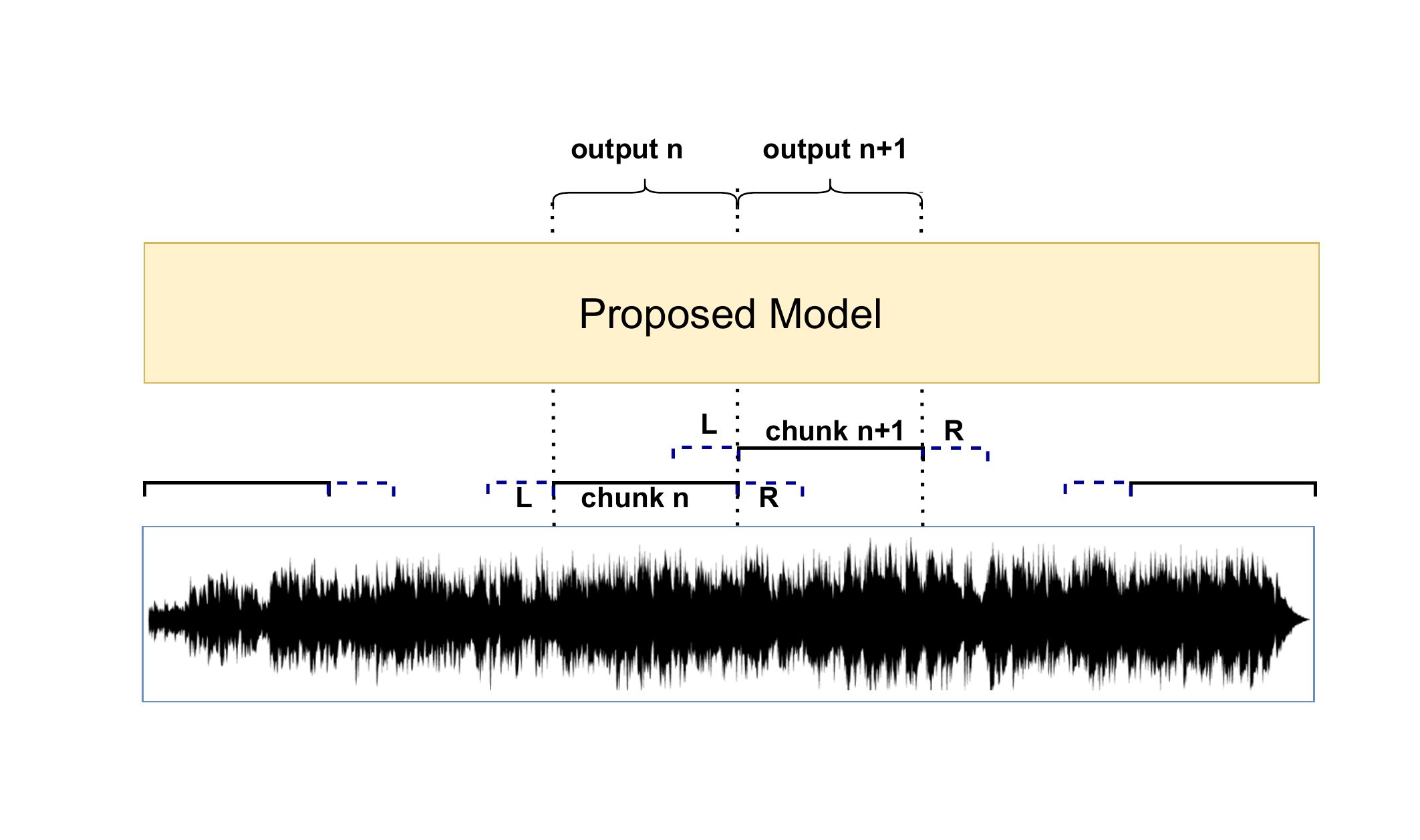}
	\caption{Chunk hopping mechanism.}
	\label{fig:chunk-hopping}
\end{figure}

To support online speech recognition, a chunk-hopping mechanism proposed in \cite{Dong2019SelfattentionAA} is implemented. The chunk-hopping mechanism is illustrated in Figure~\ref{fig:chunk-hopping}. Specifically, the complete utterance of a sentence is firstly segmented into several non-overlapping chunks. To apply contextual information, for each chunk, we splice a left chunk $L$ before the chunk as historical context and a right chunk $R$ after it as future context. Spliced chunks only serve as contexts and generate no output. For the first chunk in the utterence, only a right chunk is spliced. And for the last chunk in the utterence, only a left chunk is spliced. 

\subsection{Multi-Task Learning}
\label{sec:mtl}
For ASR task, we use a CTC loss \cite{Graves2006ConnectionistTC,Graves2014TowardsES} to optimize the model. CTC predicts the posterior probability of ${p(Y|X)}$, where ${Y}=\{y_l\in\mathcal{V}|l=1,...,L\}$ is the output sequence and ${X}=\{x_t\in\mathbb{R}|t=1,...,T\}$ is the input sequence, by introducing a framewise alignment ${A} = \{a_t\in\mathcal{V}\cup\{<b>\}|t=1,...,T \}$ with an additional blank symbol ${<b>}$. The joint probability of $Y$ given $X$ can be written as follows:
\begin{equation}
p_{CTC}(Y|X) = \sum_Ap(Y|A)p(A|X)
\label{eq1}
\end{equation}
The learning objective of a CTC-based model is defined as follows:
\begin{equation}
\mathcal{L}_{CTC} = log\ p_{CTC}(Y|X) 
\label{eq2}
\end{equation}
For VAD task, we use cross-entropy as the loss function, which is defined as follows:
\begin{equation}
\mathcal{L}_{CE} = yln\hat{y} + (1-y)ln(1-\hat{y})
\label{eq3}
\end{equation}
where y is the ground truth of VAD task; $\hat{y}$ is the prediction of VAD task.

In this paper, we use MTL to jointly train ASR and VAD tasks.  The joint loss, $\mathcal{L}_{MTL}$, is given as:
\begin{equation}
\mathcal{L}_{MTL} = \mathcal{L}_{ctc} + \mathcal{L}_{CE}
\label{eq4}
\end{equation}

\subsection{Online VAD\&ASR Inference Algorithm}
\label{sec:online-infer}
In the inference stage, to support online speech recognition and achieve robust performace, we propose an online VAD\&ASR Inference algorithm which supports ASR in a pseudo-streaming form. The online VAD\&ASR Inference algorithm is shown in Algorithm~\ref{alg:online vad/asr}. The inputs and variables that need to be initialized include:

\begin{itemize}
	\item The streaming audio input: $\textbf{x}=(x_1,...,x_t,...,x_T)$
	\item The threshold of VAD above which the current frame is classified as speech: $\theta_{vad}$ 
	\item The threshold of minimum speech length: $C$
	\item The threshold of minimum silence length: $B$
	\item The ASR chunk $\Omega_l$, which is implemented as a queue of maxsize $l$ that receives speech data, is initialized as empty queue
	\item The count of current frames that need to be processed, $c$, is initialized as 0
	\item The count of current continuous frames that are silience, $b$, is initialized as 0
	\item The current status $\mathbb{S}$, which denotes speaking if set to True and non-speaking if set to False, is initialized as False
\end{itemize}

\begin{algorithm}[!h]
	\caption{Online VAD\&ASR . }
	\label{alg:online vad/asr}
	\textbf{Input:}  $\textbf{x}=(x_1,...,x_t,...,x_T)$, $\theta_{vad}$, $C$, $B$, $l$ ; \\
	\textbf{Initialize:}  $\Omega_l\gets\emptyset$, $c=0$, $b=0$, $\mathbb{S}$ = FALSE ;\\
	\textbf{Output:}  $\textbf{y}=(y_1,...,y_i,...,y_S)$ ;
	\begin{algorithmic}[1]
		\FOR{$t=1\ ...\ T$}
		\STATE  $\theta_t$ = VAD($x_t$)
		\STATE $c$$++$
		\IF{$\theta_t>=\theta_{vad}$}
		\STATE $b=0$
		\ELSE
		\STATE $b$$++$
		\ENDIF			
		\STATE $\mathbb{S}$ = TRUE \textbf{if} $c-b>=C$
		\STATE $\mathbb{S}$ = FALSE \textbf{if} $b>=B$
		\IF{$(c-b>=l) \ ||\  ((b>=B) \&\& \mathbb{S})$}
		\STATE ENQUEUE($\Omega_l$,$(x_{t-b-c}, ..., x_{t-b})$)
		\STATE $y_i$=ASR($\Omega_l$)
		\STATE $\Omega_l\gets\emptyset$
		\STATE $c=0$
		\ENDIF
		\ENDFOR
		\RETURN $\textbf{y}=(y_1,...,y_i,...,y_S)$
		
	\end{algorithmic}
\end{algorithm}

During the online VAD\&ASR process, the proposed system continuously outputs $y_i$, which is the recognition result of ASR chunk $\Omega_l$. In lines 2-8, we predict the VAD score and make count for $c$ \& $b$ at current frame. In lines 9-10, the current status $\mathbb{S}$ is determined by $c$ \& $b$: if $c-b>=C$, which means that there are more than $C$ frames before the continuous silence frames, $\mathbb{S}$ is set to True. And if $b>=B$, which means that there are more than $B$ continuous silence frames at current time , $\mathbb{S}$ is set to False. In lines 11-16, the middle result is computed at current time $t$ if the following conditions are met: \\
\uppercase\expandafter{\romannumeral1. } $c-b>=l$, which means that there are more than $l$ frames before the continuous silence frames. \\
\uppercase\expandafter{\romannumeral2. } $(b>=B) \&\& \mathbb{S}$, which implies the end of an utterence. \\
Everytime we get the ASR result from $\Omega_l$, $\Omega_l$ and $c$ are reset to $\emptyset$ and $0$ respectively.

\section{Experimental Setup}
\subsection{Training Strategy}
\label{sec:training-strategy}
We use a two-stage training strategy to train the model. To guarantee the performance of the primary task, i.e. ASR, we train a single task ASR model by finetuning the wav2vec 2.0 pre-trained model in the first stage. In the second stage, we train our multi-task model by finetuning the ASR model from the first stage. The chunk-hopping mechanism described in Section~\ref{sec:chunk-hop} is used in the second training stage. To adapt to different real-time demands, the sizes of chunks are randomly set from 0.5 second to 3 seconds and the spliced chunks are all fixed to 0.5 second.

To obtain the labels of VAD, we use the  silero-vad\cite{SileroVAD} toolkit to generate annotations from speech.

\subsection{Dataset and Configuration}
\label{sec:configs}
The proposed system is experimented with the fairseq toolkit \cite{Ott2019fairseqAF} on the HKUST Mandarin Chinese conversational telephone speech recognition (HKUST)\cite{Liu2006HKUSTMTSAV} corpus and the Librispeech\cite{Panayotov2015LibrispeechAA} dataset. 
Librispeech dataset contains 1000h of training data split into "clean-100h", "clean-360h" and "other-500h" subsets and 
also contains development and test sets that are split into simple (“clean”) and harder (“other”) subsets. To reduce the experiment cost, we only train the model with the "clean-100h" part of the training data.
HKUST consists of long conversations with speech and non-speech parts. To recognize the long-form audios, non-speech parts need to be removed before speech parts are sent to ASR.  Therefore, HKUST dataset can be used to examine the overall effect of VAD and ASR.  As for the ASR modeling units, Chinese characters are used for HKUST dataset and letters are used for Librispeech dataset. 

We use publicly released pre-trained wav2vec2.0 base model, which is composed of a seven-block CNN feature extracter and a 12-layer transformer encoder. 
In the finetuning stage, we follow the settings described in \cite{Baevski2020wav2vec2A},  i.e. Adam optimizer, learning rate of 2e-5, total batchsize of 1600sec and a tri-stage rate scheduler where the learning rate is warmed up for the first 8000 steps, held constant for the next 32000 steps and then linearly decayed for 40000 steps.   Settings are the same for both HKUST dataset and Librispeech 100h dataset. 

In the inference stage, hyperparameters for VAD segmentation and ASR decoding are set by tuning on the development set. For HKUST, we randomly select 1000 utterances from the original training set as the development set. We use the beamsearch decoder of \cite{Pratap2019Wav2LetterAF} for ASR decoding. The LM weight and the word insertion score of a 4-gram language model (LM), which is trained on the transcript of the training set, are set to 0.46 and 0.52 respectively. The beamsize is set to 20. The threshold of VAD, $\theta_{vad}$, is set to 0.45. The threshold of minimum speech length, $C$, is set to 0.1 second. The threshold of minimum silence length, $B$, is set to 0.6 second. The size of spliced chunks described in Section~\ref{sec:chunk-hop} is set to 0.64 second.
For Librispeech, we only experiment on segmented audios and simply use the Verterbi decoder without an language model for ASR.
All the experiments are performed on 4 GeForce RTX 2080 Ti GPUs.

\section{Experimental Results}

\subsection{VAD\&ASR Multi-task Learning}
In this section, we aim to investigate the effect of VAD\&ASR Multi-task Learning (MTL) on each task. Firstly, to verify the effect of MTL on ASR task, on HKUST dataset we compare the proposed MTL method with the vanilla wav2vec 2.0 Single-task Learning (STL) system and a high-performance Transformer+CTC system on the ESPnet toolkit \cite{watanabe2018espnet}, and on Librispeech we only compare the proposed method with the vanilla wav2vec2.0 model. The results on HKUST dataset are shown in Table~\ref{tab:hkust-segment}. Character error rate (CER) is used as the evaluation metric because CER is widely used for the Chinese ASR evaluation due to its ambiguous word boundary. The results on Librispeech dataset are shown in Table~\ref{tab:librispeech}.

As can be seen from Table~\ref{tab:hkust-segment}, the baseline wav2vec 2.0 STL system achieves a very high performance on HKUST dataset: it works better than the Transformer+CTC ASR system, where the CER is reduced from 23.5\% to 22.0\%. And with the aid from the proposed MTL approach, the perfomance further improves by 7.3\% relative CER (from 22.0\% to 20.4\%).

And as shown in Table~\ref{tab:librispeech}, compared with the wav2vec 2.0 model trained on the clean 100 hour subset of Libripseech, the proposed method achieves WER 5.59/13.39 on dev-clean/other with a relative reduction of 8.7\%/1.3\% and achieves WER 5.65/12.87 on test-clean/other with a relative WER reduction of 6.9\%/3.7\%.

\begin{table}[t]
	\caption{CER(\%) comparsion on HKUST dataset}
	\label{tab:hkust-segment}
	\centering
	\begin{tabular}{l|l}
		\toprule
		{System}      & {CER}                \\
		\midrule
		Transformer+CTC (speed perturb+RNNLM)                    & 23.5                                       \\
		\midrule
		Wav2vec2.0 + 4-gram LM                  & 22.0                                \\ 
		\midrule
		Proposed + 4-gram LM              & 20.4                         \\ 
		\bottomrule
	\end{tabular}
\end{table}

\begin{table}[t]
	\caption{WER(\%) comparsion on Librispeech-100h dataset}
	\label{tab:librispeech}
	\centering
	\begin{tabular}{l|l|l|l|l}
		\toprule
		\multicolumn{1}{c|}{\multirow{2}{*}{model}} & \multicolumn{2}{c|}{dev}           & \multicolumn{2}{c}{test}          \\ \cline{2-5} 
		
		    & {clean}    & {other} & {clean} & {other}            \\
		\midrule
		Wav2vec2.0 w/o LM                 & 6.12   & 13.56 & 6.07 & 13.36                             \\ 
		\midrule
		Porposed w/o LM             & 5.59       & 13.39 & 5.65 & 12.87                        \\ 
		
		\bottomrule
	\end{tabular}
\end{table}

\begin{table}[t]
	\caption{VAD performance comparsion on HKUST dataset}
	\label{tab:hkust-vad}
	\centering
	\begin{tabular}{l|l l l}
		\toprule
		{System}      & {DetER(\%)}    & {FA(\%)}  & {Miss(\%)}             \\
		\midrule
		STL                    & 23.3           &  \textbf{4.7}  & 18.6                             \\
		\midrule
		MTL             &  \textbf{17.8}      & 5.3  &  \textbf{12.5}                    \\ 
		
		\bottomrule
	\end{tabular}
\end{table}

Furthermore, to explore performance of the proposed method on VAD task, we compare the proposed MTL method with the VAD system trained with the same CNN architecture in the proposed model architecture. We use three metrics to evaluate the performance of VAD: detection error rate (DetER), false alarm rate(FA) and missed detection rate(Miss). The DetER measures the fraction of time that is not attributed correctly to speech or to non-speech and is computed as:
\begin{equation}
DetER = \frac{N_{false\ alarm} + N_{miss}}{N_{total}}
\label{eq5}
\end{equation}
where $N_{false\ alarm}$ is the number of false positive speech predictions, $N_{miss}$ is the number of false negtive speech predictions and $N_{total}$ is the total number of predictions. 

As is shown in Table~\ref{tab:hkust-vad}, the false alarm of the proposed MTL approach is slightly higher than the STL approach. In terms of the detection error rate and missed detection, the MTL method outperforms the STL method by 23.6\% and 32.8\% relative improvements respectively.

\subsection{Online VAD\&ASR Inference}
In this section, to explore the combined effect of VAD and ASR, all expeiments are conducted on unsegmented long audios from HKUST dataset. Firstly, to investigate the performance of the random chunk-hopping training strategy decribed in Section~\ref{sec:training-strategy}, we set the ASR chunk to different lengths in the proposed online VAD\&ASR inference process. 

Results are shown in Table~\ref{tab:asr_chunk_size}, where $L_{asr}$ refers to the lengh of the ASR chunk. Obviously, the performance gets better as $L_{asr}$ increases just as it should be. The point is that it shows robustness to $L_{asr}$ in some range. When we decrease $L_{asr}$ from 5 seconds to 3 senconds, the CER only increases relatively by 1\%. But when we decrease $L_{asr}$ from 3 seconds to 0.64 sencond, the CER increase relatively by 9\%. 

\begin{table}[H]
	\caption{Effect of ASR chunk size}
	\label{tab:asr_chunk_size}
	\centering
	\begin{tabular}{l|c c c c}
		\toprule
		\textbf{$L_{asr} (s)$}      & 0.64      & 1      & 3      & 5              \\
		\midrule
		{CER} (\%)                 & 23.0               & 22.6  & 21.1 & 20.9                \\ 

		\bottomrule
	\end{tabular}
\end{table}

\begin{table}[H]
	\caption{Performance comparsion on HKUST long audio}
	\label{tab:hkust-long}
	\centering
	\begin{tabular}{l|l|l l l}
		\toprule
		{Method}      & {CER} (\%)      & {Sub} (\%)      & {Del} (\%)      & {Ins} (\%)                \\
		\midrule
		Oracle                    & 20.4           & 13.4     & 5.2     & 1.8                                 \\
		\midrule
		Base1                   & 22.6         & 14.3     & 6.6     & 1.7                           \\ 
		Base2                   & 22.0           & 14.2     & 6.0     & 1.8                          \\ 
		Proposed                  & 20.9              & 13.3  & 5.8 & 1.8                       \\ 
		\bottomrule
	\end{tabular}
\end{table}

Next, we compare the proposed system with the follwing methods:
\begin{itemize}
	\item \textbf{Oracle:} The audio input was segmented according to the manual annotations provided by the dataset
	\item \textbf{Base1:} The audio input was segmented by the GMM-based VAD system, which is implemented on the WebRTC-vad toolkit\footnote{https://github.com/wiseman/py-webrtcvad}
	\item \textbf{Base2:} The audio input was segmented by the domain-adversarial DNN-based VAD system\cite{Lavechin2020EndtoEndDV}, which is robust to domain mismatch. We use the publicly realeased pipeline implemented on the pyannote.audio toolkit\cite{Bredin2020}.
\end{itemize}

As is shown in Table~\ref{tab:hkust-long}, \textbf{Base2} outperforms \textbf{Base1}, which indicates the effectiveness of DNN-based VAD method. And the proposed system further outperforms \textbf{Base2}, where the improvment mainly comes from the decrease of substitution error. The possible reason might be that the proposed MTL method helps the model leverage linguistic information provided by ASR task. There is only a 0.25\% CER gap between the proposed method and the \textbf{Oracle} method.

\section{Conclusions}

This study proposed a novel end-to-end online ASR framework that integrate VAD by multi-task learning. The CTC-based ASR and VAD are complementary to each other so that the proposed method improved the performance on both tasks. Moreover, the proposed method used simple architecture in the bottom layers of the network to train VAD task, which resulted in a lower computational cost of VAD. The whole system was trained based on the wav2vec2.0 self-supervied pre-training method, thus reduced the reliance on labeled training data.
Our experimental reasults showed the advantages of the proposed method over other conventional methods that implement ASR after VAD segmentation.  Future work includes combining more tasks into the system and improving the inference speed.

\bibliographystyle{IEEEtran}

\bibliography{mybib}


\end{document}